# Semitransparent perovskite solar cells with an evaporated ultra-thin perovskite absorber

*Zongbao Zhang, Ran Ji, Xiangkun Jia, Shu-Jen Wang, Marielle Deconinck, Elena Siliavka, Yana Vaynzof\**

Z. Zhang, R. Ji, M. Deconinck, E. Siliavka, Prof. Y. Vaynzof
  1. Chair for Emerging Electronic Technologies, Technical University of Dresden, Nöthnitzer Str. 61, 01187 Dresden, Germany
  2. Leibniz-Institute for Solid State and Materials Research Dresden, Helmholtzstraße 20, 01069 Dresden, Germany
     E-mail: yana.vaynzof@tu-dresden.de

Dr. X. Jia, Dr. S. Wang

Dresden Integrated Center for Applied Physics and Photonic Materials (IAPP) and Institute for Applied Physics, TU Dresden, Nöthnitzerstraße 61, 01187 Dresden, Germany

Dr. X. Jia

Department of Chemical and Biomolecular Engineering, National University of Singapore, Singapore 117585, Singapore

**Abstract**:

Metal halide perovskites are of great interest for application in semitransparent solar cells due to their tunable bandgap and high performance. However, fabricating high-efficiency perovskite semitransparent devices with high average visible transmittance (AVT) is challenging because of their high absorption coefficient. Here, we adopt a co-evaporation process to fabricate ultrathin $CsPbI_3$ perovskite films. Due to the smooth surface and orientated crystal growth of the evaporated perovskite films, we are able to achieve 10 nm thin films with compact and continuous morphology without pinholes. When integrated into a p-i-n device structure of glass/ITO/PTAA/perovskite/PCBM/BCP/Al/Ag with an optimized transparent electrode, these ultrathin layers result in an impressive open-circuit voltage ($V_{OC}$) of 1.08 V and a fill factor (FF) of 80%. Consequently, a power conversion efficiency of 3.6% with an AVT above 50% is demonstrated, achieved in the 10 nm semitransparent perovskite solar cells, which is the first report for a perovskite device of 10 nm active layer with higher $V_{OC}$, FF and AVT. These findings demonstrate that evaporation process is a possible way for compact ultrathin perovskite film, which has the potential for future smart windows, light emitting diodes, and tandem device applications.

## 1. Introduction

Perovskite solar cells (PSCs) have made significant progress in recent years, reaching a record power conversion efficiency (PCE) of 26.0%, making them a promising contender to compete with commercially established silicon cells.[1–7] Due to their easily tunable bandgap and other advantageous optoelectronic properties,[8] perovskite materials are of great interest for application in semitransparent solar cells (ST-PSCs).[9,10] In contrast to conventional opaque devices, which are used on roofs or solar parks, ST-PSCs are used extensively in applications such as smart windows and displays, building-integrated photovoltaics and wearable electronics.[11] In certain cases, ST-PSCs can also be used in tandem and multi-terminal solar cell applications to make full use of the solar spectrum and obtain high-efficiency solar cells.[12–15]

Balancing device transmittance and efficiency is an essential factor in the development of ST-PSCs. In order to simultaneously improve the light transmittance and efficiency of the device, each layer of the device structure needs to be carefully designed and optimized. As the most absorbing layers of the solar cell are the perovskite active layer and the top metal electrode, their optimization drew the most attention.[16,17] Generally, there are three main methods to adjust the perovskite active layer: lowering the active layer thickness[18–20], altering its microstructure [21–23], and the use of a wide bandgap perovskite composition[24–26]. Among them, the simplest approach is to decrease the perovskite layer thickness, which in the case of solution-processed perovskites, can be achieved by reducing the precursor solution concentration. However, it has been reported that reducing the precursor solution concentration can also negatively impact on the microstructure of the perovskite layer, with a significant decrease in the grain size, and consequently a reduced open-circuit voltage ($V_{OC}$) and fill factor (FF).[27] For example, Gaspera et al. prepared a 54 nm thick perovskite film by reducing the concentration of the perovskite precursor and obtained devices with a maximum efficiency of 4.6%. [19] However, as the concentration of precursors is further reduced, obtaining compact and continuous films becomes challenging, thus resulting in a strong further reduction of the photovoltaic performance.

Deposition by thermal evaporation is a commonly used technique for the fabrication of organic and perovskite semiconductor layers and optoelectronic devices, such as solar cells, light-emitting diodes, and other devices.[28] Compared to processing from solution, deposition by evaporation is a solvent-free method, [29] which results in a slowing down of the crystal growth process, thus promoting the generation of many crystal nuclei. During the crystal growth process, the grain size is limited, which makes it possible to prepare thinner films.[30,31] Theoretically, ultrathin, compact perovskite films of various thicknesses can be obtained by an evaporation method, however, experimentally, this proved challenging since the film-forming process and the resultant crystal size need to be considered.[31] For example, by varying the deposition time during the co-evaporation process, Parrott et al. obtained perovskite films of different thicknesses, showing an obvious quantum confinement effect.[31] Paetzold and colleagues have investigated the impact of the substrate by depositing thin perovskite films on different materials. In all cases, the authors observed that the ultrathin perovskite films (7 nm and 20 nm) exhibited poor morphology with many pinholes.[30] Bolink and coworkers fabricated compact perovskite films of 50 nm and 100 nm by thermal evaporation, reaching an efficiency of more than 9% and a light transmittance of 23%.[32] These approaches suggest that a minimum thickness of ~50 nm is required for the fabrication of a continuous and compact perovskite layer, regardless of whether solution processing or thermal evaporation is used, thus posing a limit to the maximum average visible transmittance (AVT) that can be achieved.

To further increase the AVT of ST-PSCs, there is a need to understand the limiting factors in the formation of ultra-thin films (<50 nm) and develop strategies to mitigate them. Importantly, ultrathin films are also of great interest for application in light-emitting diodes and other electronic devices due to the combination of quantum confinement effects and good charge transport.[31] Despite this, there are relatively few studies examining the formation of ultrathin perovskite films and photovoltaic devices based on such layers have not been reported.[33–37]

Herein, we investigate how the growth mechanism of thermally evaporated perovskites impacts the microstructure of ultrathin films. We demonstrate that perovskite layers that grow in a highly oriented fashion make it possible to prepare ordered, compact and smooth ultrathin perovskite films as thin as 10 nm. When integrated into a solar cells architecture with an optimized semitransparent electrode, the 10 nm thick layers achieved excellent photovoltaic performance with a FF of 80% and $V_{OC}$ of 1.08 V, resulting in an efficiency of 3.6% with an AVT of 54.26%. To the best of our knowledge, this is first demonstrated of efficient ST-PSCs of 10 nm active layer with AVT>50%. Our work provides valuable insights for the development of high-efficiency ultra-transparent ST-PSCs required for their future integration into applications such as smart windows and tandem solar cells.

## 2. Results and Discussion

### 2.1. Growth of ultrathin perovskite film

To explore the possibility of producing ultrathin perovskite films and identify the possible role of the growth mechanism in this process, we selected two different perovskite compositions: an inorganic perovskite, $CsPbI_3$, and an organic-inorganic perovskite $Cs_{0.1}FA_xPbI_{2+x}Br_{0.1}$. Based on our previous work, adding a small amount of phenethylammonium iodide (PEAI) as an additive can significantly improve the quality and stability of the inorganic $CsPbI_3$ film formed by thermal evaporation.[38] Therefore, three sources were used for the deposition of the perovskite layers: $PbI_2$, CsI, and PEAI for $CsPbI_3$ and $PbI_2$, CsBr, and FAI for $Cs_{0.1}FA_xPbI_{2+x}Br_{0.1}$ (**Figure 1**a). In the following, we denote the inorganic $CsPbI_3$ film with a PEAI additive as PEACs and the hybrid $Cs_{0.1}FA_xPbI_{2+x}Br_{0.1}$ perovskite as FACs. The composition of these two perovskites were probed by X-ray photoemission spectroscopy (XPS) and XPS depth profiling and is summarized in Table S1. We note that the composition of thin films is not stoichiometric and is different from that of thicker films,[38] which has been observed for thermally evaporated perovskite layers in the past.[39] Detailed fabrication procedures for each of the films are provided in the experimental section.

To monitor the changes in the film morphology of the two different perovskite compositions with increasing film thickness, the layers were characterized by scanning electron microscopy (SEM) and atomic force microscopy (AFM). Considering that the growth of the perovskite layers is impacted by the choice of underlying substrate,[30] we utilized ITO/PTAA substrates that would be later used for the fabrication of ST-PSCs. In both cases, the microstructure of the films consists of small grains, as is common for thermally evaporated perovskites (Figure 1c-h).[29] Interestingly, regardless of films thickness, the PEACs films exhibit a smooth surface, evidenced by a root mean square (RMS) roughness value of ~2.5nm, which is very close to the roughness of the ITO/PTAA substrates (1.7 nm, Figure S1). On the other hand, the surface of FACs perovskite films is significantly rougher (RMS roughness of ~7 nm), similar to the observations previously made for inorganic-organic perovskites in literature.[32] For the 10 nm layer thickness, the PEACs film is compact, homogeneous, and smooth, whereas the FACs layer exhibits a discontinuous morphology with many pinholes. To explore whether PEACs perovskite can also be deposited in thinner, compact layers, we deposited 2 nm and 5 nm films. SEM images (Figure S2) revealed the presence of many islands on the surface of 2 nm PEACs films, which begin to form a more continuous film as more material is deposited. This is evidenced by the 5 nm thick film, which still exhibits some pinholes, which are eliminated entirely for the 10 nm thickness. This observation suggests that the perovskite grows via a Volmer-Weber type growth mode, which is consistent with the previously reported results.[31] Consequently, for the PEACs perovskite composition, the minimum thickness for a continuous film is approximately 10 nm, which is close to the perovskite grain size. On the other hand, depositing 10 nm thick FACs films does not lead to the formation of a continuous compact layer, and even a 20 nm thick film was found to exhibit pinholes (Figure S3). These experiments suggest that the PEACs composition has a substantially lower limit for the formation of a continuous film. To explore whether the perovskite compositions (organic-inorganic hybrid perovskite versus inorganic perovskite) induce these different limits, we also characterized additive-free $CsPbI_3$ perovskite films (i.e., without PEAI) of 10 nm and 100 nm thickness. As can be seen in Figure S4, 10 nm thick additive-free $CsPbI_3$ films are discontinuous, and only for far thicker films (100 nm), a typical small grain microstructure is observed.[40] This suggests that the different thickness limitation is not only impacted by the composition of the deposited perovskite layer but also by the growth mechanism, which will be explained in the following. Interestingly, the difference between the two perovskite layers can also be observed by examining their optical properties. Figure 1b displays the absorbance spectra of 100 nm PEACs and FACs films. The spectra reveal that the PEACs film has a sharper absorption edge than FACs, implying improved optoelectronic quality of the PEACs perovskite layers.[41] We note that the FACs film exhibits stronger absorption beyond 680 nm as a result of its narrower band gap (Figure S5).[42,43]

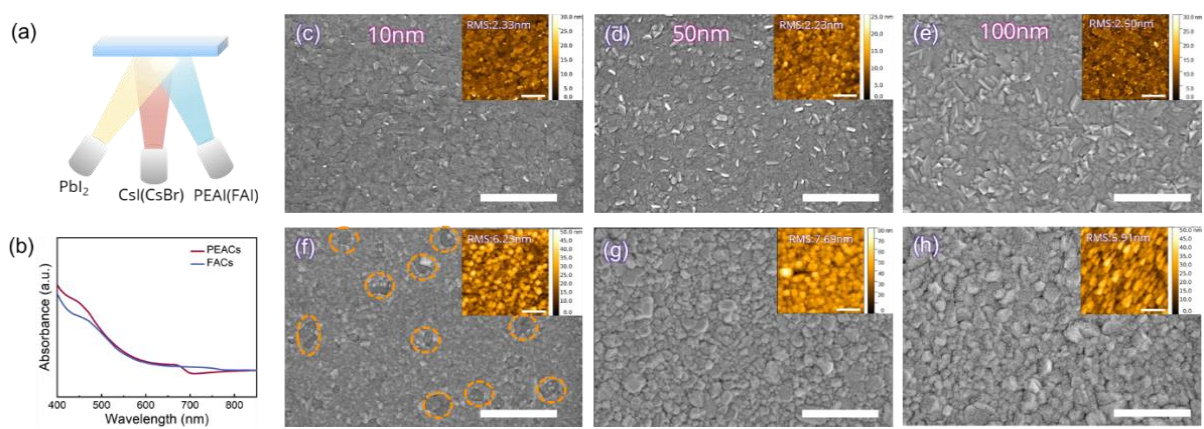

**Figure 1.** a) Schematic illustration of perovskite film preparation, b) Absorbance spectra of 100 nm perovskite films with two kinds of perovskite (PEACs and FACs). SEM images of different kinds of perovskite layer with different thickness, PEACs: c) 10 nm, d) 50 nm and e) 100 nm; FACs: f) 10 nm, g) 50 nm, and h) 100 nm. Inset figure is the corresponding AFM images. For all figures, the scale bar is 500 nm.

To further investigate the properties of the thin films, their crystallinity was examined by X-ray diffraction (XRD) with the corresponding diffraction patterns shown in **Figure 2**. Even in the case of the ultra-thin perovskite films (10 nm), the diffractograms of both perovskite compositions exhibit the typical characteristic peaks at approximately 14° and 28.5°, without the presence of any additional impurities such as $PbI_2$, suggesting that both perovskites are formed with good crystallinity and high purity by thermal evaporation.[30,31,39,44] As the thickness of the perovskite films increases, the main perovskite diffraction peaks become sharper and stronger and peaks associated with other crystal orientations become more evident. The PEACs films display the preferred (00l) crystal orientation, whereas FACs films exhibit preferred the (h00) orientation. To better quantify the degree of preferred orientation, we calculated the ratios between the peak intensities of the preferred plane to other crystalline planes (Figure S6). The results reveal that as compared to the FACs perovskite films, the PEACs are significantly more aligned along the (00l) planes, which is consistent with earlier findings showing that the use of larger organic A cation molecules can induce a preferred (00l) orientation.[45–47] The (00l) planes are parallel to the substrate, implying that perovskite crystals grow along the c-axis, which is known to result in improved optoelectronic performance and good charge transport in perovskite devices compared to other crystal facets.[48–50]

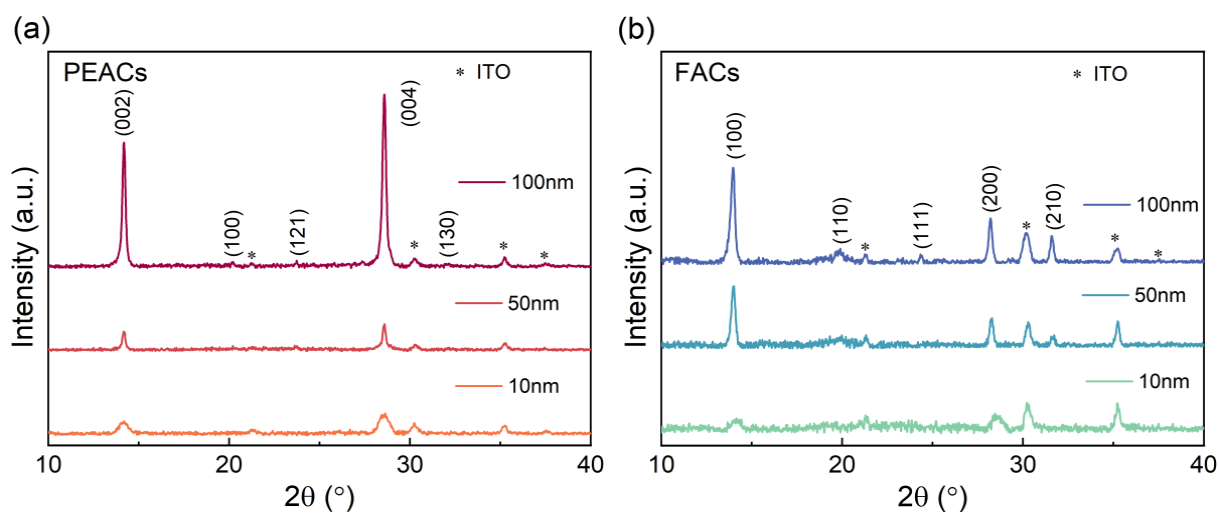

**Figure 2**. X-ray diffraction (XRD) patterns of perovskite film on ITO/PTAA substrates with different thicknesses: a) PEACs perovskite film, b) FACs perovskite film.

Taken together, the XRD and SEM results allow us to propose a basic model of the crystallization process in the ultrathin perovskite films based on the two compositions we explore (**Figure 3**). We propose that there are two different scenarios for the growth of the co-evaporated perovskite: (1) a columnar growth mode and (b) a disordered growth mode. Due to the preferential growth along the (00l) plans, the PEACs perovskite tends to adopt the columnar mode, as illustrated in Figure 3a. This crystal growth mode can help form a uniform and smooth perovskite film and enables the formation of compact ultrathin layers.[30,49] On the other hand, the growth of the FACs perovskite is not as strongly oriented, so it follows a more disorderly growth mode (Figure 3b). The difference between the two films is also evident in the cross-sectional SEM images of the 100 nm thick films (Figure 3c and 3d), where the FACs films exhibit significantly more boundaries in the vertical direction, while the PEACs result in a much more homogeneous grain structure. Notably, the FACs film contains voids at the interface between the perovskite and the ITO/PTAA substrates as well as in the bulk of perovskite layer. These voids are further evidence for the disordered growth mode and are consistent with the previous observations of pinholes and high surface roughness of the ultrathin FACs films, both of which are expected to negatively impact on the device performance.[49]

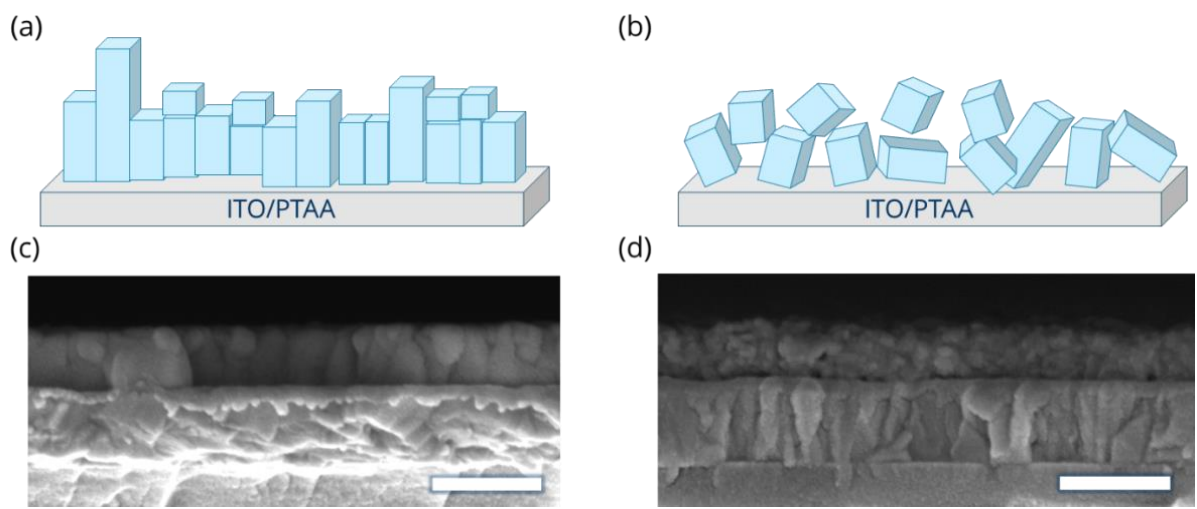

**Figure 3.** Basic crystallization model for two kinds of perovskite composition on ITO/PTAA substrates: a) PEACs deposited on the ITO/PTAA substrates, b) FACs deposited on ITO/PTAA substrates. Cross-sectional SEM images of 100 nm perovskite layers on the ITO/PTAA substrates: c) PEACs; d) FACs. For all SEM images, the scale bar is 200 nm.

## 2.2. Optical properties and photovoltaic performance

To investigate the suitability of the PEACs and FACs perovskite films for application in semitransparent solar cells, we first characterized their optical properties. **Figure 4**a shows the transmittance spectra of the PEACs and FACs perovskite film with different thicknesses. For the 10 nm perovskite film, the FACs shows a much higher transmittance in the entire 350-800 nm range as compared to the PEACs perovskite, which is a consequence of the many pinholes in the 10 nm FACs perovskite (see Figure 1d). As the thickness of the perovskite layer increases, the ITO/PTAA substrates are fully covered and the transmittance of the perovskite films decreases. For thicker films such as 50 nm and 100 nm, the PEACs films exhibit much higher transmittance than their FACs counterparts, which is related to the larger bandgap of the former. To quantify the differences in transmittance, the average visible transmission (AVT) is used to evaluate the optical transmission of perovskite film in the visible spectrum. The AVT is defined as follows[51,52]:

$$\text{AVT} = \frac{\int T(\lambda)P(\lambda)S(\lambda)d\lambda}{\int P(\lambda)S(\lambda)d\lambda}$$

Where the λ is the wavelength range of 390-780 nm, T is the transmittance of measured sample, P is the photopic response of human eyes, which is shown in Figure 4a. S is the solar photon flux intensity defined by the air mass (AM) 1.5 spectra. The AVT values for both series of the perovskite samples are summarized in Table S2, ranging from ~37% to ~79%. Another crucial factor for semi-transparent devices is color neutrality. A semi-transparent film with a neutral color (i.e., close to the white region) is more suitable for applications such as smart windows. In general, the color neutrality can be assessed by the color coordinate of the transmittance light in the international commission on illumination (CIE) 1931 color space. Figure 4b demonstrates that ultrathin perovskite films (10 nm) exhibit more neutral coloration close to white region, which indicates considerable potential for producing high-quality semi-transparent perovskite devices.[52,53] This color neutrality can also be observed by eye when examining text placed behind the PEACs and FACs samples (Figure 4c). Furthermore, the steady-state and time-resolved photoluminescence (TRPL) measurements were performed to examine the quality of perovskite films and corresponding results are shown in Figure S7-8 and Table S3. For both perovskite compositions, a strong blue shift is observed with a decrease in the film thickness (Figure S7), which originates from the quantum confinement in the thin films, described by Herz et al. in their previous work.[31] Figure S8 and Table S3 display the TRPL decay curves for the two different perovskite compositions and varying film thicknesses. Compared to the inorganic PEACs perovskite, FACs perovskites show a longer PL lifetime, which is consistent with previous reports.[38,54–56] Moreover, we observe that the PL lifetime is reduced with decreasing the thickness of perovskite films in both perovskite compositions.

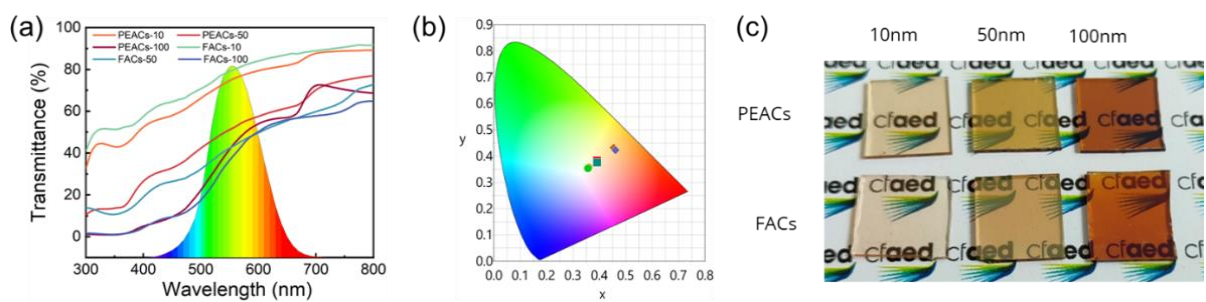

**Figure 4.** a) The human luminosity curve and transmittance spectra of PEACs and FACs perovskite films with different thickness; b) Color coordinates plotted on the CIE 1932 chromaticaity diagram and c) Corresponding photos of semi-transparent PEACs and FACs perovskite films with different thicknesses.

To investigate the photovoltaic performance of the different perovskite samples, they were integrated into a p-i-n device structure with ITO/PTAA/perovskite/PCBM/BCP/Ag architecture, as is shown in **Figure 5**a. As a starting point, these devices include an opaque Ag electrode (80 nm). The current-density-voltage (J-V) curves of the best devices and the distribution of the photovoltaic parameters are displayed in Figure 5b-f. Generally, the PEACs devices lead to much higher $V_{OC}$s and FFs than those of the FACs. The enhanced $V_{OC}$ is at least partly attributed to the larger bandgap of the PEACs, however the fact that some 100 nm PEACs devices reach a $V_{OC}$ of 1.18 V, also suggests that the optoelectronic quality of the thermally evaporated perovskite is very high. In fact, to the best of our knowledge, this is one of the highest $V_{OC}$ values previously reported for evaporated perovskite devices without any surface treatment.[54,57–60] In comparison to the PEACs perovskite devices of the same thickness, devices based on FACs show much broader photovoltaic parameter distributions and significantly lower FF values. Interestingly, considering the much lower bandgap of the FACs perovskites, an increased short-circuit current ($J_{SC}$) would be expected in comparison to that of the PEACs devices, yet Figure 5e shows that the $J_{SC}$ of both types of devices are very similar. Figure 5b shows the JV curves of the champion cells with the associated photovoltaic parameters summarized in Table S4. Figure S9 shows the EQE spectra and the integrated $J_{SC}$ for champion devices, with the latter being consistent with the values in the JV curves. The performance of the 10 nm FACs device is particularly poor, consistent with our observations of a poor morphology of perovskite active layer. To the contrary, the 10 nm PEACs perovskite devices show decent performance.

To further examine the role of PEAI in the formation of ultrathin perovskite films, solution-processed $CsPbI_3$ films of various thicknesses were prepared by utilizing PEAI as an additive. For consistency, the same amount of PEAI was used as an additive in solution-processed layers as that in the optimized evaporated layers (Figure S10-12, Table S5). The thickness of the layers was varied by changing the concentration of the perovskite precursor solution. As shown in Figure S13, films with higher concentration (0.4 M) resulted in films with a complete coverage and a thickness of approximately 160 nm. Lowering the concentration to 0.2 M results in thinner layers (~65 nm) and is accompanied by the emergence of pinholes. As the concentration is further decreases to 0.1 M, the films become thinner (~40 nm) and the number of pinholes markedly increases. For the lowest concentration of 0.05M, the films consist of many islands with a corresponding thickness is approximately 25 nm. These results are consistent with previous reports which demonstrate that pinholes and islanding are commonly observed for ultrathin solution-processed perovskite films.[19,20,27] The layers were incorporated into photovoltaic devices, whose photovoltaic performance parameters distributions are summarized in Figure S14 and the performance of champion devices are shown in Figure S15 and Table S6. In comparison to the evaporated PEACs perovskite devices, solution-processed devices show notably lower $V_{OC}$ and FF values as is expected for thin films with a discontinuous morphology. Taken together, the SEM and photovoltaic performance results suggest that the impact of PEAI in the formation of thermally evaporated $CsPbI_3$ films cannot be translated to their processing from solution, and thus cannot be used to reduce the thickness limitation for the formation of compact solution processed perovskite layers.

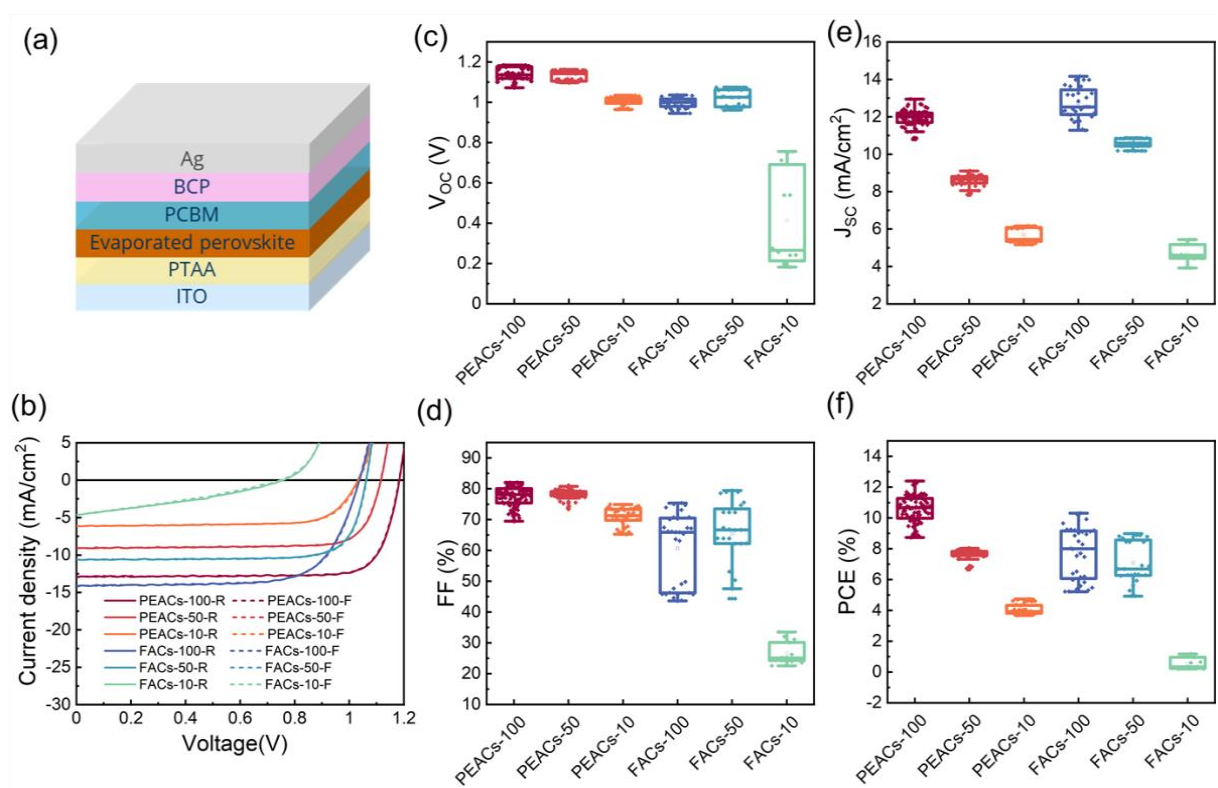

**Figure 5**. a) Structure of devices in this work. b) JV curves of PEACs and FACs perovskite devices with various thicknesses and 80 nm Ag electrode. Phototvoltaic performance parameters: c) $V_{OC}$, d) FF, e) $J_{SC}$ and f) PCE distributions of PEACs and FACs perovskite devices with various thicknesses. A total of 257 devices were measured.

## 2.3. Performance of ultra-thin perovskite films with optimal transparent electrode

Another critical factor in determining the transmittance of ST-PSCs is the top electrode. To select the optimal transparent electrode, we first integrated a 100 nm thick PEACs perovskite in devices with a variety of electrode types. These include a thick Ag (80 nm), thin Ag (7 nm or 10 nm) with Al seed layer (0.5 nm), an ITO electrode, and ITO with an Al doped ZnO (AZO) buffer layer. The device performance and transmittance spectra are shown in Figure S16 and Table S7. While devices with ITO-based electrodes exhibit higher transmittance, their photovoltaic performance is significantly reduced, in particular in the $V_{OC}$ and FF. The insertion of an AZO buffer layer between the PCBM and ITO slightly mitigates this issue, but the overall performance is still rather low.

Reducing the thickness of the metal contact is a well-established method to increase transparency without significantly sacrificing the device's performance. The use of ultrathin metal electrodes is typically accompanied by the use of buffer layers,[61] fabrication of dielectric-metal-dielectric multilayer layers,[19,62] or insertion of a seed layer.[32,63] We utilized the latter approach and deposited an ultrathin seed layer of 0.5 nm Al in order to test thin Ag layers of either 7 nm or 10 nm in thickness. The optimal interplay between high transmittance and high photovoltaic efficiency was obtained for a 10 nm Ag layer, which was selected for integration into semitransparent solar cells with perovskite active layers of different thicknesses. The photovoltaic performance parameter distributions of devices with PEACs (thickness 10 nm, 50 nm and 100 nm) and FACs (50 nm) with 10 nm thick Ag electrodes are summarized in **Figure 6**a-d. Compared to PEACs devices of any thickness, the FACs devices result in a lower $V_{OC}$ ($\approx$ 1.0 V) and FF ($\approx$ 68%), which is attributed to their rough surface, the presence of voids and the disordered crystal orientation. On the other hand, the PEACs devices exhibit higher $V_{OC}$ values, although a notable reduction is observed for decreasing active layer thickness. Still, the ultrathin devices with a 10 nm perovskite active layer reach $V_{OC}$ as high as 1.08 V. Remarkably, the reduction in thickness has no negative impact on the FF of the PEACs devices, which remains at approximately 80%. The $J_{SC}$ is reduced with decreasing active layer thickness, as is expected due to the reduced light absorption.

Figure 6e displays the JV curves for the corresponding champion cells of each type with the photovoltaic parameters listed in **Table 1**. We note that in addition to their lower performance, the FACs devices exhibit obvious hysteresis. Figure 6f displays the EQE spectra and corresponding integrated $J_{SC}$ for the champion cells, which match well with the $J_{SC}$ values extracted from the JV curves. Furthermore, the EQE spectra and integrated $J_{SC}$ are in good agreement with the results of optical device simulations (Figure S17-18, Table S8). Figure 6g shows the transmittance spectra of all devices and corresponding color coordinate plots on CIE 1932 chromaticaity diagram are shown in Figure 6h. As can also be seen from the inset of Figure 6g, the PEACs devices with a 10 nm thick perovskite active layer and 10 nm thick electrode exhibit high transmittance and good color neutrality, making them particularly promising for a potential application in smart windows or other applications with significant needs for high AVT.

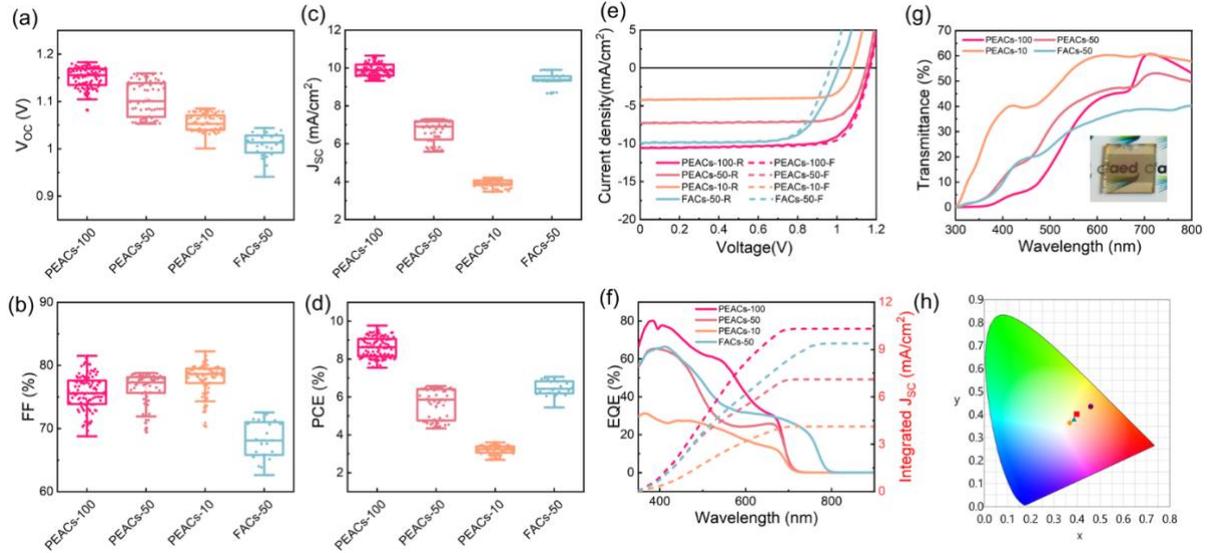

**Figure 6.** Photovoltaic performance parameters: a) $V_{OC}$, b) FF, c) $J_{SC}$ and d) PCE distributions and corresponding e) JV curves of PEACs and FACs perovskite devices with different thickness. A total of 273 devices were measured. f) EQE spectra, g) transmittance spectra and h) Color coordinates plotted on the CIE 1932 chromaticity diagram of PEACs perovskite with different thickness perovskite active layer.

**Table 1.** Photovoltaic performance parameters of ST-PSCs on different perovskite composition with various thicknesses

| Device | Scan direction | Voc [V] | Jsc [mA/cm²] | FF [%] | PCE [%] | AVT(%) |
|---|---|---|---|---|---|---|
| **PEACs-100** | Forward | 1.17 | 10.65 | 78.09 | 9.77 | 31.89 |
| | Reverse | 1.17 | 10.65 | 75.62 | 9.40 | |
| **PEACs-50** | Forward | 1.16 | 7.21 | 78.53 | 6.58 | 39.98 |
| | Reverse | 1.16 | 7.21 | 78.33 | 6.58 | |
| **PEACs-10** | Forward | 1.08 | 4.19 | 79.74 | 3.60 | 54.26 |
| | Reverse | 1.08 | 4.19 | 79.68 | 3.60 | |
| **FACs-50** | Forward | 0.97 | 9.90 | 72.55 | 6.93 | 30.62 |
| | Reverse | 1.01 | 9.90 | 70.73 | 7.07 | |

Finally, we compared the photovoltaic performance of the PEACs ultrathin devices with those previously reported for thin perovskite active layers (≤ 200 nm). As is shown in **Figure 7**, generally, a reduction in the active layer thickness is accompanied by a strong decrease in the $V_{OC}$ and FF of the devices, with very few exceptions. The PEACs perovskite devices, on the other hand, exhibit very high $V_{OC}$ (1.1 V) and FF (80%) in comparison to previously reported solution processed and thermally evaporated semitransparent perovskite devices.[19,64–80] Due to the ultrathin perovskite active layer, the transmittance of the 10 nm PEACs device is significantly increased, reaching an AVT above 50%. As is shown in Figure 6c, this device is the first demonstration of an efficient ST-PSC with such high AVT.[19,20,22,61,62,64,81–89]

The stability of ST-PSCs is another important factor for their future application in commercial devices. Previous reports suggested that ultrathin perovskite layers are fragile and may rapidly decompose.[30,31] Figure 7d displays the shelf-storage stability tests of 10 nm thick PEACs devices over 35 days, demonstrating that the devices maintain their efficiency upon storage. The operational stability (under continuous illumination) of the ultrathin unencapsulated devices is reduced in comparison to that of devices with thicker perovskite active layers (Figure S19, Figure S20),[38] as a consequence of the sensitivity of thin perovskite films to oxygen and humidity.[30,39] While these observations highlight the importance of developing surface passivation layers, more stable extraction layers and encapsulation strategies for enhancing the stability of semitransparent perovskite solar cells, this goes beyond the scope of this work.

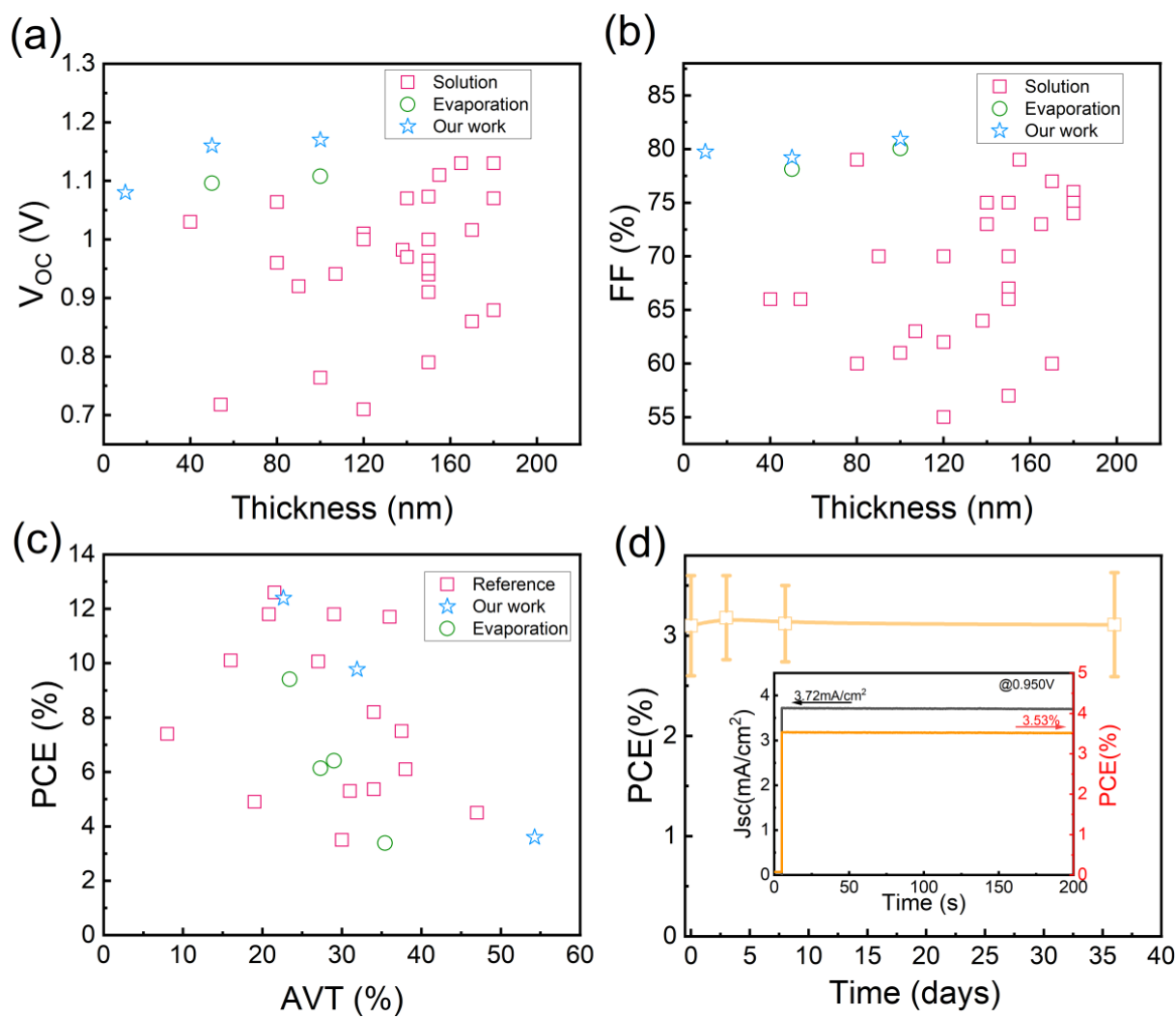

**Figure 7.** Plots of a) $V_{OC}$ and b) FF as a function of thickness for comparison with other semitransparent PSCs (thickness ≤ 200nm). c) PCE as a function of AVT for comparison with other semitransparent PSCs. d) PCE stability characteristics of unencapsulated 10nm ST-PSCs with thin Ag electrode. Inset figure is the MPP tracking of 10 nm semitransparent devices with thin Ag electrode.

## 3. Conclusion

To summarize, we demonstrated that it is possible to form ultrathin (10 nm), continuous, and smooth perovskite films using the thermal evaporation processes. The introduction of small amounts of PEAI during the deposition of $CsPbI_3$ films enables to control the crystal growth and its orientation, leading to the formation of compact and smooth thin films with thicknesses below 50 nm. Other perovskite compositions, such as FACs or $CsPbI_3$ deposited without the PEAI additive do not result in compact ultrathin films as their growth is not highly oriented, but is rather disordered. When integrated into photovoltaic devices with optimized transparent electrodes, the ultrathin PEACs perovskite layers reach an efficiency of 3.6% with a high AVT of 54.26% - the highest reported to date for perovskite semitransparent PSCs of AVT over 50%. These findings demonstrate the importance of controlling the orientation of thermally evaporated perovskite layers, thus enabling a columnar growth mode, required for the formation of compact and homogenous ultrathin layers, which is promising for future applications in smart windows, light-emitting diodes and tandem devices.

## 4. Experimental Section

*Materials:* ITO substrates were obtained from PsiOTech Ltd. $PC_{61}BM$ (>99.5%) was bought from Luminescence Technology Crop. PTAA was purchased from Sigma-Aldrich. Lead (II) iodide (99.99%, trace metals basis) was obtained from TCI company. Cesium iodide (99.999%, metals basis) and formamidinium iodide (FAI) were bought from Sigma-Aldrich. Aluminum doped ZnO (AZO) nanoparticle dispersion in alcohol solvent (N-20X-Flex) were obtained from Avantama Company. Phenylethylammonium iodide (PEAI) was obtained from Greatcellsolar Materials. Chlorobenzene, Toluene, Isopropanol were bought from Acros Organics. All materials were used without more purification.

*Perovskite Film Deposition:* Clean substrates were transferred to a vacuum chamber (CrePhys GmbH, Germany) from a nitrogen-filled glovebox. For PEA-$CsPbI_3$ and $Cs_{0.1}FA_xPbI_{2+x}Br_{0.1}$ perovskite, they were prepared in two separated chambers to avoid the contamination. When the pressure was pumped down to around $10^{-6}$ mbar, the evaporation process started to form individual perovskite composition. The deposition rates and layer thickness were monitored using calibrated quartz crystal microbalances (QCM, Rate/thickness monitor STM-100/MF, Syncon Instrument). The tooling factor of evaporated materials adopted two ways. For inorganic composition, they were tooled by comparing the thickness signal of QCMs close to the crucibles to the thickness of film deposited on clean glass substrates. The thickness of films on substrates was measured with Dektak profilometer. Due to the soft property of organic components such as PEAI and FAI, measuring thickness of these organic materials on glass was not possible. So a QCM balance was mounted at the sample position and the tooling factor was obtained via comparing the signals of two QCM balances. For the PEA-CsPbI$_3$ perovskite evaporation, the three sources (PbI$_2$, CsI and PEAI) were preheated to achieve a desirable rate ≈0.6–0.8 Å s$^{-1}$ (310–330 °C) for PbI$_2$, 0.65–0.75 Å s$^{-1}$ (445–455 °C) for CsI, 0.001–0.144 Å s$^{-1}$ (120–145 °C) for PEAI. For all evaporated PEA-CsPbI$_3$ film, the rate ratio of PbI2 and CsI was fixed to a constant value of 1:0.85, corresponding a molar ratio of 1: 1.11. And the volume concentration of PEAI to PbI$_2$ was fixed to 5%. For Cs$_{0.1}$FA$_x$PbI$_{2+x}$Br$_{0.1}$ evaporation, the whole evaporation process were controlled by software SweepMe! (https://sweep-me.net). The rates of FAI, CsBr and PbI$_2$ were fixed at 1.50 Å s$^{-1}$ (120-130 °C), 0.06 Å s$^{-1}$ (385-395 °C), and 0.83 Å s$^{-1}$ (290-300 °C) respectively. For thickness of all perovskite films, they were measured with Dektak profilometer.

*Device fabrication:* Patterned ITO was rinsed with acetone to remove the protective glue. Then they were ultrasonically cleaned with 2 % hellmanex detergent, deionized water, acetone, and isopropanol for 20 minutes respectively, followed dry with nitrogen gun. For PTAA hole transport layer, the ITO substrates was treated with oxygen plasma for 10 min and then spin-coated with PTAA (1.5mg/mL in toluene) at 6000 rpm 30 s, followed by annealing at 100 °C for 10 min in a nitrogen filled glovebox. Subsequently, HTLs coated substrates were transferred to vacuum chamber for evaporation. By detecting the rates of different sources, the perovskite layers with different thicknesses were fabricated. Next, 25 μL PC$_{61}$BM solutions (20mg/mL in chlorobenzene) was dynamically spin-coated at 2000 rpm for 30 s followed by annealing at 100 °C for 3 min. Finally, 35 μL hole-blocking layer BCP (0.5mg/mL in isopropanol) was dynamically deposited on substrates at 4000 rpm for 30 s, following by 80 nm thermally evaporated Ag cathode (Mantis evaporator, base pressure of 10$^{-7}$ mbar). For thin metal electrode evaporation, the samples are transferred to another nitrogen-filled metal chamber. Firstly, a 0.5 nm Al seed layer were evaporated on the perovskite film, followed by 7 or 10 nm Ag deposition. For ITO electrode sputtering, prior to depositing electrode, 30 uL AZO nanoparticle dispersion solution dynamic spin-coated on the substrates, followed by sputtering 500 nm ITO (Quorum, Q150T ES).

*Photovoltaic Device Characterization*: EQE spectra of the devices were recorded using the monochromatic light of a halogen lamp from 400 nm to 800 nm, the reference spectra were calibrated using the NIST-traceable Si diode (Thorlabs). J-V characteristics were recorded by using a computer controlled Keithley 2450 source measure unit under a solar simulator (Abet Sun 3000 Class AAA solar simulator). The incident light intensity was calibrated via a Si reference cell (NIST traceable, VLSI) and tuned by measuring the spectral mismatch factor between a real solar spectrum, the spectral response of reference cell and perovskite devices. All devices were scanned from short circuit to forward bias (1.3 V) to and reverse with a rate of 0.025 V s$^{-1}$. No treatment was applied prior to measurements. The active area for all devices was 4.5 mm$^2$.

*UV-vis Absorption and photoluminescence measurements:* A Shimadzu UV-3100 spectrometer was utilized to record the ultraviolet−visible (UV−vis) absorbance spectra. PL measurements were performed using a CW blue laser (405 nm, 10 mW, Coherent) as the excitation source. The PL signal was collected using a NIR spectrometer (OceanOptics). All samples were with encapsulated to prevent the decomposition and enable the all measurements to be were carried out in ambient air at room temperature.

*Time-Correlated Single Photon Counting (TCSPC):* A TCSPC setup contained of a 375 nm laser diode head (Pico Quant LDHDC375), a PMA Hybrid Detector (PMA Hybrid 40), a TimeHarp platine (all PicoQuant), and a Monochromator SpectraPro HRS-300 (Princeton Instruments). Perovskite films on quartz were excited with the 375 nm laser diode and then the emission was collected by the PMA hybrid detector. The pulse width is ≈44 ps, power is ≈3 mW, the spot size is ≈1 mm$^2$, so the excitation fluence is ≈0.132 J m$^{-2}$. The lifetimes were evaluated using reconvolution algorithms of FluoFit (PicoQuant).

*Scanning-Electron Microscopy (SEM):* A SEM (Gemini 500, (ZEISS, Oberkochen, Germany)) with an acceleration voltage of 3 kV was utilized to obtain the surface and cross-sectional morphology images.

*Atomic Force Microscopy (AFM):* AFM measurements were performed with a Dimension ICON3 scanning probe microscope from Bruker AXS S.S.S under ambient conditions in the ScanAsyst mode in air using RTESPA-150 tips. The size of the record images was 2×2 μm$^2$ and the scan rate ranged between 0.5-1.0 Hz at 1024 points per line. The samples used for AFM imaging were perovskite layers deposited on ITO/PTAA substrates.

*X-ray diffraction (XRD)*: XRD patterns were measured in ambient air by using a Bruker Advance D8 diffractometer equipped with a 1.6 kW Cu-Anode (λ = 1.54060 Å) and a LYNXEYE_XE_T 1D-Mode detector. The scans (2theta-Omega mode, 2θ = 10°-40°, step size 0.01°, 0.1 s/step) were measured in Standard Bragg-Brentano Geometry (goniometer radius 420 mm) with a height limiting slit of 0.2 mm.

*X-ray photoemission Spectroscopy (XPS) measurement:* The samples were transferred to an ultrahigh vacuum chamber (ESCALAB 250Xi by Thermo Scientific, base pressure: 2 × 10−10 mbar) for XPS measurements. XPS measurements were carried out using an XR6 monochromated Al Kα source (hν= 1486.6 eV) and a pass energy of 20 eV. Depth profiling was performed using an argon gas cluster ion beam with large argon clusters (Ar2000) and an energy of 4 k eV generated by a MAGCIS dual mode ion source. During XPS depth profiling, the etching spot size was (2.5 × 2.5) mm$^2$ and the XPS measurement spot size was 650 μm. The measurement time per etch level was 8 min.

**Supporting Information**

Supporting Information is available from the Wiley Online Library or from the author.

**Acknowledgements**

Z. Z. and R.J. are grateful for the financial support by the China Scholarship Council (Scholarship#201806750012 and #201806070145, respectively). Z.Z thanks short-term scholarship from the Graduate Academy of Technische Universität Dresden and the Dresden Center for Nanoanalysis (DCN) for providing access to the SEM measurement. Z.Z appreciates Prof. Alexey Chernikov and Prof. Karl Leo for providing the facility for AFM measurement and transparent electrode evaporation, respectively. Z.Z also thanks Dr. Martin Kroll, Anna-Lena Hofmann, Shaoling Bai, Dr. Juanzi Shi, Dr. Rongjuan Huang, Dr. Fabian Paulus, and Sophia Terres for their insightful and fruitful discussions. S.-J.W. acknowledges funding from DFG Project, WA 4719/2-1 and support from the Hector Fellow Academy. This project has received funding from the European Research Council (ERC) under the European Union's Horizon 2020 research and innovation programme (ERC Grant Agreement n° 714067, ENERGYMAPS) and the Deutsche Forschungsgemeinschaft (DFG) in the framework of the Special Priority Program (SPP 2196) project PERFECT PVs (#424216076).